\documentclass[pre,twocolumn,amsmath,amssymb,floatfix]{revtex4}

\usepackage{graphicx}% Include figure files
\usepackage{dcolumn} % Align table columns on decimal point
\usepackage{bm}      % Bold math
\usepackage{color}

\begin{document}

\title{Effect of helicity and rotation on the free decay of turbulent flows}
\author{T. Teitelbaum$^{1}$ and P.D. Mininni $^{1,2}$}
\affiliation{$^1$ Departamento de F\'\i sica, Facultad de Ciencias Exactas y
         Naturales, Universidad de Buenos Aires, Ciudad Universitaria, 1428
         Buenos Aires, Argentina. \\
             $^2$ NCAR, P.O. Box 3000, Boulder, Colorado 80307-3000, U.S.A.}
\date{\today}

\begin{abstract}
The self-similar decay of energy in a turbulent flow is studied in 
direct numerical simulations with and without rotation. Two initial 
conditions are considered: one non-helical (mirror-symmetric), and 
one with maximal helicity. The results show that, while in the absence 
of rotation the energy in the helical and non-helical cases decays with 
the same rate, in rotating flows the helicity content has a major impact on 
the decay rate. These differences are associated with differences in the 
energy and helicity cascades when rotation is present. Properties of the structures that arise in the flow at late times in each time are also discussed.
\end{abstract}
\maketitle

%\section{Introduction}

Turbulence is ubiquitous in nature, and many turbulent flows are also 
rotating. The effect of rotation becomes important when the Rossby number 
(the ratio of the convective to the Coriolis acceleration) is sufficiently 
small. Mid-latitude synoptic scales in the atmosphere, stellar convective 
regions, and turbomachinery are examples of such flows. Helicity (alignment of the velocity and the vorticity) is also important 
for many processes in astrophysical, geophysical and engineering flows. 
As an example, helical flows were proposed as the reason for the 
stability of rotating convective thunderstorms \cite{Lilly86}.

Studies of isotropic and homogeneous helical turbulence 
\cite{group1,Kraichnan73} showed that both the helicity 
and the energy are transferred toward smaller scales with 
constant fluxes. Moreover, it was observed that the scaling of the 
energy (Kolmogorov's law) was unchanged by the presence of helicity. 
As a result, helicity is expected to globally arrest the 
energy decay, but not to change its self-similar decay rate. Non-helical 
rotating turbulence has been studied in detail (see e.g., 
\cite{group2} and references therein), but perhaps because 
of the similarities between the helical and non-helical cases in 
non-rotating flows, not much attention has been paid to helical rotating 
turbulence.

The lack of detailed studies of rotating helical flows is remarkable 
considering the relevance of both helicity and rotation in many 
astrophysical and geophysical processes. In this work, we study the 
effect of rotation and of helicity in the self-similar decay of energy 
in turbulent flows. Even in isotropic and homogeneous turbulence, the 
law for the decay rate of energy is a matter of debate \cite{group5}. 
It is known that it depends on properties of the infrared energy spectrum 
(i.e., the spectrum at scales larger than the energy containing scale), 
and may depend on other statistical properties of the initial conditions. 
As a result, in this letter we will consider only two flows with the same 
infrarred spectrum and the same energy decay rate in the absence of 
rotation, and study how the presence of helicity and rotation changes 
their decay. New decay laws are found for helical rotating flows in numerical simulations.
The results are then interpreted in terms of how the 
energy and helicity cascades are modified when rotation is present, 
and a phenomenological theory that is in agreement with the simulations 
is finally discussed.

%\section{Numerical simulations}

The numerical simulations solve the Navier-Stokes equations for an 
incompressible fluid in a rotating frame,
\begin{equation}
\partial_t {\bf u} + \mbox{\boldmath $\omega$} \times
    {\bf u} + 2 \mbox{\boldmath $\Omega$} \times {\bf u}  =
    - \nabla {\cal P} + \nu \nabla^2 {\bf u} ,
\label{eq:momentum}
\end{equation}
where ${\bf u}$ is the velocity field ($\nabla \cdot {\bf u} =0$), 
$\mbox{\boldmath $\omega$} = \nabla \times {\bf u}$ is the vorticity, 
${\cal P}$ is the total pressure (modified by the centrifugal term), 
and $\nu$ is the kinematic viscosity. We chose the rotation axis to be 
in the $z$ direction, $\mbox{\boldmath $\Omega$} = \Omega \hat{z}$, 
with $\Omega$ the rotation frequency. Our integration domain is a periodic 
box of length $2\pi$. Two sets of runs were done at resolutions of $256^3$ 
(set A) and $512^3$ grid points (set B) using a pseudo spectral code. The parameters for all the runs 
are listed in Table \ref{table:runs}.

\begin{table}
\caption{\label{table:runs}Parameters used in the simulations: $\nu$ is 
the kinematic viscosity, $\Omega$ is the rotation rate, $Re$ is the 
Reynolds number, $Ro$ is the Rossby number, $Ro^{\omega}$ is the 
micro-Rossby number, $E_k$ is the Ekman number, and $h$ is the relative 
helicity of the initial conditions. The values of $Re$, $Ro$, $Ro^{\omega}$, 
and $E_k$ are given at the time of the peak of dissipation $t^*$.}
\begin{ruledtabular}
\begin{tabular}{ccccccccc}
Run&$\nu$&$\Omega$&$Re$&$Ro$&$Ro^\omega$&$E_k$ &$h$  \\
\hline
A1 &$1.5\times 10^{-3}$& $0$ & $450$  & $-$ & $-$ &
  $-$ &$0$ \\
A2 &$1.5\times 10^{-3}$& $0$ & $600$  & $-$ & $-$ &
  $-$ &$0.95$ \\
A3 &$1.5\times 10^{-3}$& $4$ & $550$  & $0.12$ & $1.28$ &
  $2.2\times 10^{-4}$ &$0$ \\
A4 &$1.5\times 10^{-3}$& $4$ & $830$  & $0.083$ & $0.8$ &
  $1.0\times 10^{-4}$ &$0.95$ \\
B1 &$7\times 10^{-4}$& $4$ & $1100$ & $0.12$ & $1.82$ &
  $1.1\times 10^{-4}$ &$0$ \\
%B2 &$7\times 10^{-4}$& $4$ & $?$ & $?$ & $$ &
%  $?$ &$0.4$ \\
B2 &$7\times 10^{-4}$& $4$ & $1750$ & $0.083$ & $1.15$ &
  $4.7\times 10^{-5}$ &$0.95$ \\
\end{tabular}
\end{ruledtabular}
\end{table}

To simulate systems with different amount of relative initial helicity 
($h = H/\left<|{\bf u}||\mbox{\boldmath $\omega$}|\right>$ where 
$H=\left<{\bf u}\cdot\mbox{\boldmath $\omega$}\right>$ is the flow net 
helicity), two flows were considered as initial conditions: the 
Taylor-Green (TG) flow \cite{Taylor37}, and the Arn'old-Beltrami-Childress 
(ABC) flow \cite{Childress}. The TG flow is non-helical, and has zero 
energy in the $k_z=0$ mode, whose amplification observed in the rotating 
cases (see below) is thus only due to a cascade process. The TG flow was 
chosen for its importance in hydrodynamics; it was originally motivated 
as an initial condition which leads to rapid development of small spatial scales. 
It also mimics the von K\'arm\'an flows between two counter-rotating 
disks used in several experiments. % \NOTE{of rotating turbulence \cite{Simand00}}.
The ABC flow is an eigenfunction of the curl operator and as a result 
has maximum helicity. It was used as a paradigmatic example to study helical 
flows in the atmosphere \cite{Lilly86}. Both flows develop, after a 
short time, an infrared energy spectrum proportional to $k^2$; this is 
important to ensure we can compare the decay rates of both flows.
%\NOTE{equal foot}.

The simulations were started using a superposition of these flows 
from wavenumbers $k=4$ to $14$. Runs with zero relative helicity 
have a supperposition of TG flows and runs with $h\approx 0.95$ have 
a supperposition of ABC flows. %and run B2 with $h\approx 0.4$ has a 
%mixture of both.
The initial energy spectrum from $k=4$ to $14$ was 
proportional to $k^{-4}$, leaving enough spectral space 
to allow for direct and inverse transfer of energy. All the runs were 
extended for over 40 turnover times,
and the dissipative range was properly resolved. Times in 
Table \ref{table:runs} and in the figures are expressed in units of the 
turnover time at $t=0$, $T=L/U$, where $L\approx 2\pi/k_0 = 2\pi/4$ 
is the initial integral scale of the flow and $U\approx 1$ is the initial
 r.m.s. velocity.

Several Reynolds, Rossby, and Ekman numbers can be defined for the runs. 
We will consider here the Reynolds number based on the integral scale 
$Re = UL/\nu$, and the accompanying Rossby number 
$Ro = U/(2 \Omega L)$. The integral scale is defined as 
$L = 2\pi \int{E(k) k^{-1} dk}/E$ where $E(k)$ is the isotropic energy 
spectrum and $E$ is the total energy. The Ekman number is then defined 
as $E_k=Ro/Re$. At $t=0$ runs in set A have $Re = 1047$ and runs in set 
B have $Re = 2244$. At the same time the Rossby number is $Ro = 0.08$ 
for runs A3, A4, and B1-B2; the initial Ekman number is then $E_k = 7.6 \times 10^{-5}$ 
in runs A3 and A4, and $E_k = 3.6 \times 10^{-5}$ in runs B1-B2. Values of the 
controlling parameters for each run at the time of maximum dissipation $t^*$
are given in Table \ref{table:runs}.

It is also convenient to introduce a micro-scale 
Rossby number as $Ro^{\omega} = \omega/(2 \Omega)$ \cite{Jacquin90}. 
This number can be interpreted as the ratio of the convective to the 
Coriolis acceleration at the Taylor scale, a scale characteristic of 
the turbulent inertial range. In order for anisotropies to develop in the
simulations, the Rossby number $Ro$ must be small enough for the rotation 
to affect the free decaying turbulence, but the micro-Rossby 
$Ro^{\omega}$ must be larger than one for scrambling effects of inertial 
waves not to completely damp the nonlinear terms (which leads to a pure 
exponencial viscous decay) \cite{Cambon97}. We are 
thought interested in simulations with moderate rotation rates to ensure 
$Ro \lesssim 1$ and with $Re$ large enough to have enough scale 
separation between the energy-containing and the dissipative scale. Note 
in Table \ref{table:runs} how $Ro$ and $Ro^{\omega}$ are one order of 
magnitude apart in all runs at time $t^*$. This difference 
is sustained in time in the runs during the self-similar energy decay.

%\section{\label{sec:evol}Time evolution without rotation}

\begin{figure}.
\includegraphics[width=8cm]{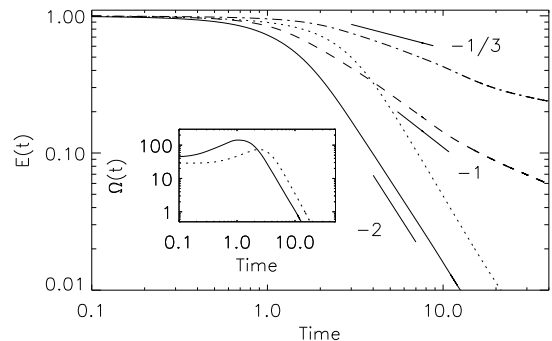}
\caption{Energy evolution for runs A1 (solid), A2 (dotted),
B1 (dashed) and B2 (dot-dashed). Runs A1 and A2 have $\Omega=0$ and decay with the same rate 
independently of the helicity content. When rotation is turned on the decay rates differ.
Inset: enstrophy evolution for runs A1 and A2.}
\label{fig:time1} \end{figure}

Figure \ref{fig:time1} shows the time history of the energy and enstrophy 
in runs A1, A2, B1-B2. In all runs there is a self-similar decay after 
the time $t^*$. Runs A3 and A4 (not shown) decay 
respectively as runs B1 and B2, although the time span of the self-similar 
stage is shorter. In both runs with $\Omega=0$, the energy decays as 
$\sim t^{-2}$. As will be discussed later, this 
exponent corresponds to the decay of a flow with constant integral length 
\cite{Biferale03}.

The presence of helicity in run A2 doesn't seem to affect the self-similar 
decay of turbulence, as predicted in Ref. \cite{Kraichnan73}. This is in 
good agreement with the fact that helicity doesn't change the spectral 
index of the energy in non-rotating turbulence \cite{group1}. 
However, the self-similar decay in run A2 starts at a later time, as also 
previously reported in Ref. \cite{Morinishi01}. This is associated with the 
slow-down in the generation of small-scales in helical flows 
\cite{Lesieur}, which results in a longer time to reach the maximum of 
dissipation in run A2 (see the inset in Fig. \ref{fig:time1}). 

%\section{\label{sec:transfer}Time evolution with rotation}

In runs with rotation (A3, A4, B1 and B2), a transient is also observed 
before $t^*$. Then, self-similar decays with different power 
laws for the energy are found in all runs. Runs A3 and B1 have a decay 
near $t^{-1}$ \cite{group3,bellet}, while the runs with maximum helicity 
(A4 and B2) follow a decay slightly faster than $t^{-1/3}$. % A run 
%with intermediate relative helicity (run B2) lies between these two 
%extreme cases.
Contrary to the non-rotating case, we can appreciate now 
how the presence of initial helicity changes the self-similar energy decay.

%{\bf The time evolution of the enstrophy for runs B1 and 
%B2 is also shown in the inset of Fig. \ref{fig:time1} where wee see again 
%how the maximum of enstrophy is reached at differents times for each flow.}
%{\bf [esta ultima oracion sacarla o dejarla pero no poner nada en le inset]}

\begin{figure}
\includegraphics[width=8cm,height=4cm]{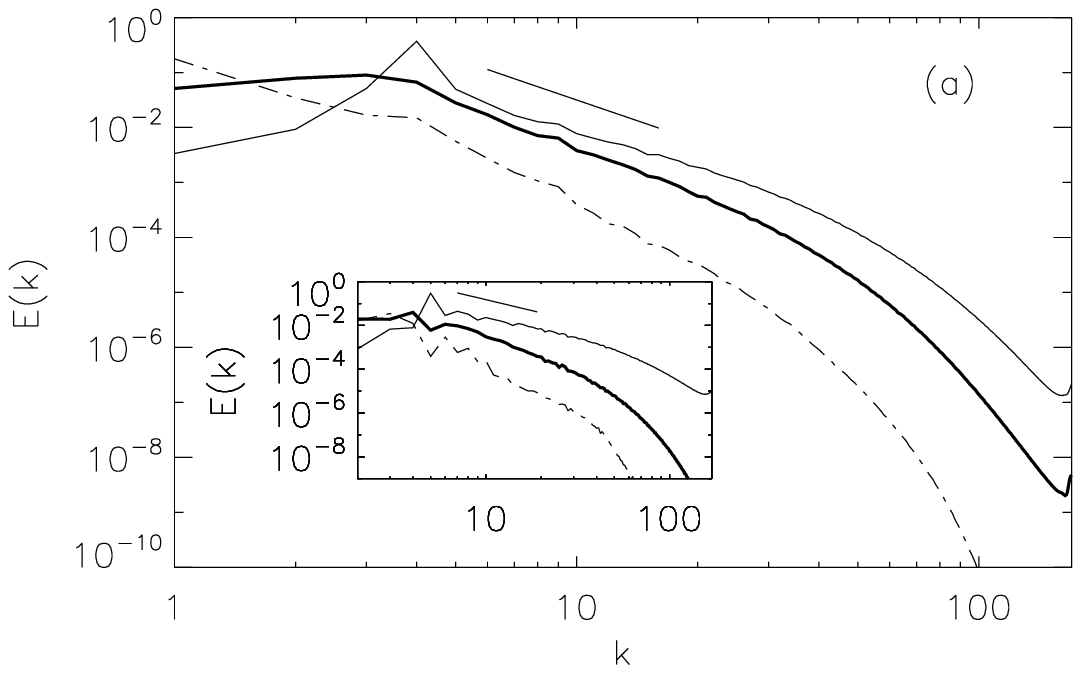}
\includegraphics[width=8cm,height=4cm]{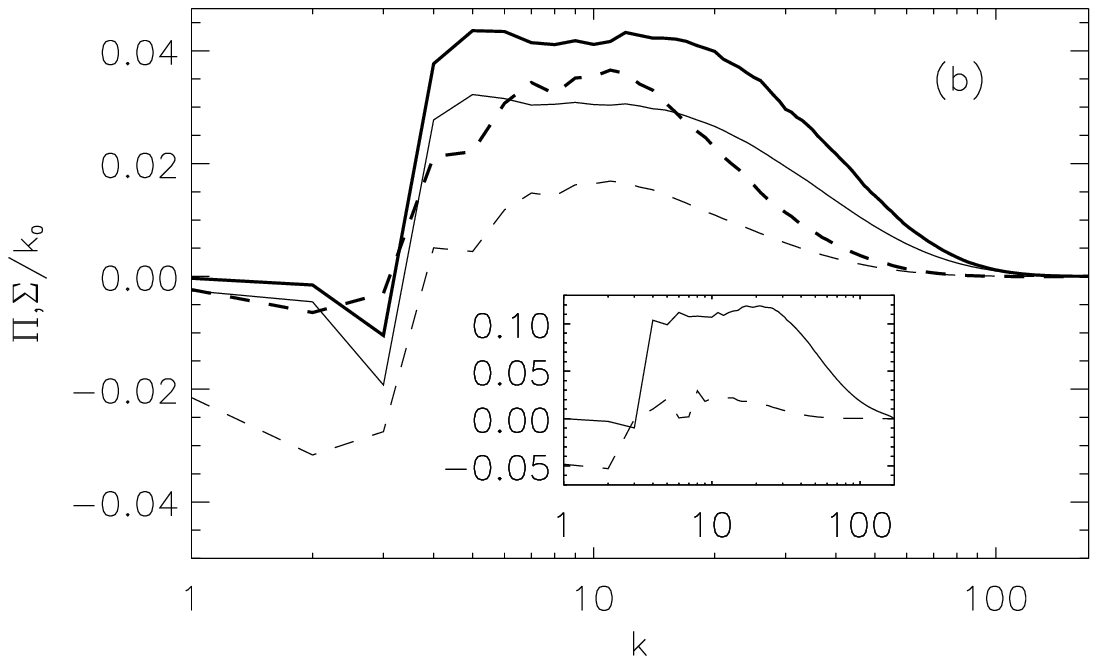}
\caption{(a) Isotropic energy spectrum for run B2 at times $t \approx 2$ (thin), $t \approx 12$ (thick) and $t \approx 30$ (dot-dashed). At early times the spectrum follows a $\sim k^{-2.5}$ power-law (slope shown as a reference) and becames later dominated by the piling up of energy in the large scales. Inset: Isotropic energy spectrum for run B2. A $k^{-2}$ slope is plotted for reference.
(b) Energy (thin) and helicity (thick) normalized fluxes for run B2 at times $t \approx 5$ (solid) and $t \approx 13$ (dashed) respectively. %Observe how a direct helicity cascade dominates at early times and the coexistence of an inverse energy and a direct helicity flux for later times.
Inset: Energy flux for run B1 at times $t \approx 2$ (solid) and $t \approx 11$ (dashed).}
\label{fig:espectros} \end{figure}

The time evolution is accompanied by a change in the shape of the energy 
spectrum (see Fig. \ref{fig:espectros}(a)). While the runs without rotation 
show at $t^*$ and during the self-similar decay an 
spectrum consistent with Kolmogorov scaling in both the energy and the 
helicity, runs with rotation show different scaling laws. Run B1 (which 
has a larger scale separation than run A3) shows an isotropic energy 
spectrum compatible with $k^{-2}$ (the anisotropic energy spectrum is 
also compatible with a $k_\perp^{-2}$ scaling) \cite{muller,MininniPress}. On the 
other hand, the energy spectrum in run B2 is slightly steeper than in 
run B1, while the helicity spectrum in this run is shallower than the 
energy spectrum. The product of both spectra is consistent with a 
$k^{-4}$ scaling law, as predicted using phenomenology for rotating 
helical flows in \cite{MininniPress}.

In the runs with rotation, a change in the small-scale spectrum is 
observed at $t \approx 20$. As energy piles up at the largest available 
scale in the box and column-like structures form in the velocity, the 
small-scale energy spectrum becomes steeper. This is accompanied by a 
decrease in the decay rate of the energy (see Fig. \ref{fig:time1}). This 
process is reminiscent of the change observed in the free-decay of 
two-dimensional turbulence, when the coalescence of large-scale vortices 
at late times leads to a steeper energy spectrum and a change in the 
self-similar decay \cite{McWilliams84} 

%In run B2 the isotropic energy spectrum also develops an early power law behavoir with exponent $-2$ ($t \sim 2$) which changes to $-2.5$ some turn over times later ($t \approx 10$) toghether with a pile up of energy at large scales as for the TG case (citas). At time $t^* \approx 20$ the largest scales dominates and a steeper spectrum is observed (not shown) ($ \approx -2.7$). The predominance of large structures is observed in both B1 and B2 \NOTE{eventualmente B2} runs and can also be evidenciated by the slow down of the energy decay observed in Figure \ref{fig:timeer} for $t>20$ (solid) and $t>10$ (dashed). In section "anisotropy and strucutres" we will find more evidence of this coalescence of ... and we will also visualize the flows and actually observe the presence of few big structures (column-shaped vortical structures) for late times. 
%Later on, the spectrum is a power law with exponent $< \approx 3$ and the largest scales are still dominant.

%{\bf Flux:}

%\begin{figure}
%\includegraphics[width=8cm]{flujos_1.eps}
%\caption{Energy (thinn) and helicity (thick) normalized fluxes for run B2 at times $t \approx 5$ (solid) and $t \approx 13$ (dashed) respectively. Observe how a direct helicity cascade dominates at early times and the coexistence of both, inverse energy and direct helicity fluxes for later times. Inset:Energy flux for run B1 at times $t \approx 1.7$ (solid) and $t \approx 11$ (dashed).}
%\label{fig:flujos} \end{figure}

Runs A1 and A2 developt direct energy fluxes toward small-scales 
while the rotating non-helical runs (A3 and B1) show both a direct and an 
inverse energy cascade (see the inset of Fig. \ref{fig:espectros} (b)).
In the helical runs (Fig. \ref{fig:espectros} (b)), at $t \approx 2$ we observe maximum flux 
of energy and helicity toward smaller scales, evidenciating both energy 
and helicity have a direct cascade. However, the helicity flux is larger 
than the direct energy flux. Later, an inverse cascade of energy can be 
clearly identified from the negative energy flux at large scales. At 
$t \approx 13$, the coexistence of both an inverse cascade of energy and 
direct cascades of energy and helicity is observed.

\begin{figure}
\includegraphics[width=8cm]{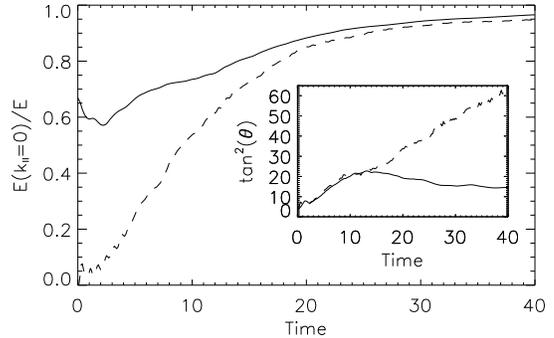}
\caption{Ratio of the energy in modes with $k_{||}=0$ to the total energy in runs B1 (dashed) and B2 (solid). Despite the different initial conditions, the curves grow monotonically to 1, showing a transfer of energy to perpendicular modes. Inset: time evolution of the Shebalin angle for the same runs.}
\label{fig:ani1} \end{figure}

%In all runs, as the time of saturation of the energy at the largest 
%available scale and of coalescence of columnar structures is reached 
%($t \approx 20$), the inverse energy flux at the largest scales 
%becomes positive bouncing back to smaller scales.

%\section{\label{sec:anis} anisotropy and structures}

%\begin{figure}
%\includegraphics[width=8cm]{ani2.eps}
%\caption{Evolution of the ratio of perpendicular to parallel integral scales for runs B1 (dashed) and B2 (solid). The %inset shows the time evolution of the parallel integral scale for the same runs.}
%\label{fig:ani2} \end{figure}

Although Figs. \ref{fig:espectros} (a) and (b) show the 
isotropic spectra and fluxes, the flows in the rotating case are 
anisotropic. Most of the energy in the spectra is in modes perpendicular 
to the axis of rotation, and the anisotropic spectum $E(k_\perp)$ and 
flux $\Pi(k_\perp)$ look similar to the ones previously discussed. 
We present instead some global indications of the development of anisotropies.
Figure \ref{fig:ani1} shows the ratio of the energy 
in perpendicular modes (i.e., modes with $k_{||}=0$) to the total energy 
in runs B1 and B2. Differences at early times are due to differemt 
initial conditions. However, it can be seen that both curves grow 
monotonically to a value near 1, indicating the flows evolve toward 
anisotropic states as the energy is transferred to perpendicular modes 
\cite{group2}.

%Note also that due to the fourier descomposition of the initial flows used (TG and ABC) $E_{\perp}/E$ starts at $0$ for B1 and at aproximately $2/3$ for B2.

A measure of anisotropies in the small-scale fluctuations is given 
by the Shebalin angle, defined as
\begin{equation}
\tan^2\theta=2 \frac {\sum_{k=1}^{k_{max}}{k_{\perp}^2E(k_{\perp})}}{\sum_{k=1}^{k_{max}}{k_{\parallel}^2E(k_{\parallel})}},
\label{eq:shebalin}
\end{equation}
The evolution of this angle in runs B1 and B2 can be seen in the inset of 
Fig. \ref{fig:ani1}. It grows monotonically although in run 
B2 it reaches a maximum at $t \approx 12$ and then seems to saturate. A 
similar behavior is observed for $L_{\parallel}$ (not shown) which again grows monotically in 
B1 but reaches a maximum in B2 at aproximately the same time. In all cases, $\tan^2\theta>>2$ which corresponds to anisotropic flows.

\begin{figure}
\includegraphics[width=8cm]{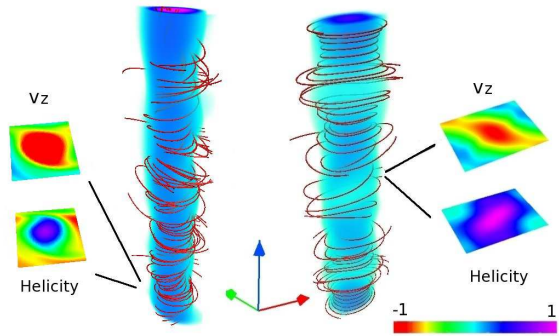}
\caption{(Color online) Visualizations of the r.m.s. vorticity for $t \approx 45$ with superimposed velocity field lines for runs A3 (right) and A4 (left). Cuts of $v_z$ and $H$ on a plane perpendicular to \mbox{\boldmath $\Omega$} are shown. The color table gives the amplitudes normalized to the maximum of each quantity.}
\label{fig:columnas} \end{figure}

The increase of the correlation lengths, together with the growth of 
$E(k_{||}=0)/E$ and of $\tan^2\theta$ speaks of anisotropization  
of the flows %as time evolves and the inverse cascade of energy develops.
This tendency towards two-dimensionalization is confirmed by exploring the flows in real space. Fig. \ref{fig:columnas} shows visualizations of the r.m.s. vorticity 
with superimposed velocity field lines for runs A3 and A4 at late times 
($t \approx 45$). In both cases, a strong anisotropy is observed 
with large scale column-like structures in the vorticity (similar 
structures are observed if energy density is visualized 
instead). However, a significant difference in the geometry of the flow in the columns 
is observed between runs A3 and A4 . While the flow in 
the columns of the helical run is strongly helical %\NOTE{h=0.9;hr=1 }
(the flow goes either up or down in the entire column, thus giving helical 
velocity field lines), the columns of the non-helical run have no net 
helicity (the flow goes up and down inside the same column). This indicates 
that even at late times, the properties of the emerging structures in 
rotating flows depend on the initial helicity content.

%\section{\label{sec:scaling}Phenomenology}

The results presented so far show a novel and distinct evolution in the 
free decay of helical and non-helical flows when rotation is present. 
These differences can be understood in terms of a simple phenomenological 
theory. For the non-rotating cases we make use of the classical Kolmogorov 
phenomenology which leads to the well known energy spectrum 
$E(k) \sim \epsilon^{2/3}k^{-5/3}$, where $\epsilon$ represents the energy 
dissipation rate. From the Navier-Stokes equation it is easy to show that for a freely-decaying flow 
$dE/dt \sim E^{3/2}/L$ which leads to a self-similar decay law 
$E(t) \sim t^{-2}$ if the integral scale is assumed constant \cite{Biferale03,group5}.

% we have seen that an energy cascade dominates the flux over the inertial range. Considering the energy flux in the inertial range $\epsilon$ is slowed down by $\epsilon \sim \frac{u_l^2}{t_l} (\frac{t_{\Omega}}{t_l}) $ where $u_l$ is the velocity at the scale $l$, $\tau_l \sim l/u_l$ is the eddy turnover time at the same scale and $\tau_{\Omega} \sim 1/\Omega$ is the characteristic time of inertial waves. This expression takes into account the slowing-down of transfer to small scales due to three-wave interactions(referencias:Iroshnikov and Kraichnan, [25]y[26] de "scale interactions and scaling laws in rotating flows at moderate..").

In the case of rotating non-helical flows, it is often assumed that 
$E(k) \sim \epsilon^{1/2}\Omega^{1/2}k^{-2}$ (for simplicity, we use the isotropic 
wavenumber $k$, although the arguments here can be easily generalized to the 
anisotropic case replacing $k$ by $k_\perp$). This spectrum was observed 
in simulations \cite{muller}, and obtained from closures \cite{bellet}. It can 
also be derived from phenomenological arguments assuming the inertial 
waves slow down the energy cascade \cite{Zhou95}. Using this spectrum it 
follows that $dE/dt \sim (E/L)^{2}$ resulting in a decay law 
$E(t) \sim t^{-1}$.

The novel case of rotating helical flows differs from the previous two cases. 
The direct transfer is dominated by the helicity cascade. Writting the 
helicity flux as $\delta \sim h_l/(\Omega \tau_l^2)$ where $h_l$ is the 
helicity at the scale $l$ and $\tau_l$ the eddie turnover time, leads 
to spectra $E(k) \sim k^{-n}$ and $H(k) \sim k^{n-4}$, where $n=5/2$ for the 
case of maximum helicity \cite{MininniPress}. In this case, 
dimensional analysis leads to $E(t) \sim \epsilon^{1/4}\Omega^{5/4}k^{-5/2}$ 
and to $E(t) \sim t^{-1/3}$. Note that for flows with initial helicity 
between zero and the maximum, the decay rate is between $-2$ and $-1/3$.
A run with relative helicity of $ h \approx 0.4$ was done to verify this.

%-> increase of the correlation lengths along the axis of rotation associated with a two-dimensionalization process (citar experimentos y simulaciones que aparecen en paper de Bourouiba & P Bartello (pag 140). Fijarme si los papers que tengo muestran esta longitud paralela).

%-> Fijarse: se forman ciclones o anti-ciclones (alineacion del vector vorticidad con el eje de rotacion. Rompimiento de simetria?leer Metais & Lesieur (1994))

%\section{Conclusions}
%- Las estructuras observadas se generan para rotacion no muy grade.
%- 2-dimensionalizacion del flujo
%- Fenomenologia se condice con las simulaciones

% It is interesting to point out that 
%although both the energy and the helicity direct cascade, the direct 
%helicity flux is expected to dominate the direct range as the Rossby 
%number $Ro$ is decreased \cite{Mininni08}. It has been observed in 
%simulations of forced rotating helical turbulence that the sum of the 
%difference between the direct and inverse energy and helicity fluxes 
%remains constant independly of $Ro$, meaning that the amount of direct 
%helicity flux is associated with the energy flux to smaller wave-lenths. 
%Increasing the rotation rate $\Omega$ more energy transfer to small 
%scales would developt and the helicity direct flux would become more 
%dominant.

The results indicate that helical and non-helical rotating flows differ in their scaling laws and decay rates. This is different than the case of non-rotating flows where helicity doesn't change the energy scaling. The observed differences indicate that helical rotating flows must be studied in detail, and open new directions in the study of rotating turbulence. This can also be the starting point to elucidate the actual role helicity plays in the decay of flows in nature where rotation is often relevant \cite{Lilly86}.

%\begin{acknowledgments}

{\it Computer time provided by NCAR and CECAR. The authors acknowledge support from grant UBACYT X468/08. PDM is a member of the Carrera del Investigador Cient\'{\i}fico of CONICET. Flow visualizations were done using VAPOR \cite{vapor}.}

%\end{acknowledgments}

%\bibliographystyle{acm}
%\bibliography{ms}

\begin{thebibliography}{10}

\bibitem{Lilly86}
{D.K. Lilly, Atm. Sc. {\bf 40}, 126 (1986).}

\bibitem{group1}
{V. Borue and S.A. Orszag, Phys. Rev. E {\bf 55}, 7005 (1997); Q. Chen, S. Chen, and G.L. Eyink, Phys. Fluids {\bf 15}, 361 (2003); P.~D. Mininni, A. Alexakis, and A. Pouquet, Phys.\ Rev.\ E {\bf74}, 016303 (2006).}

\bibitem{group2}
{C. Cambon and L. Jacquin, J. Fluid Mech. {\bf 202}, 295 (1989); F. Waleffe, Phys.\ Fluids A {\bf5}, 677 (1993); C. Cambon, N.N. Mansour, and F.S. Godeferd, J. Fluid Mech. {\bf 337}, 303 (1997).}

\bibitem{group5}
{P.G. Saffman, Phys. of Fluids {\bf 10}, 1349 (1967); U. Frisch, {\it Turbulence: The legacy of A.N. Kolmogorov} (Cambridge Univ.\ Press, Cambridge, 1995); L. Skrbek and S.R. Stalp, Phys. Fluids {\bf 12}, 1997 (2000); P.A. Davidson,{\it Turbulence} (Oxford University Press, 2004).}

\bibitem{Taylor37}
{G.I. Taylor and A.E. Green, Proc.\ Roy.\ Soc.\ Lond.\ Ser.\ A {\bf158}, 895 (1937), 499.}

\bibitem{Childress}
{S. Childress and A.D. Gilbert, ({\it Stretch, Twist, Fold: The fast dynamo}(Springer-Verlag Berlin, 1995).}

\bibitem{Jacquin90}
{L. Jacquin {\it et al}, J.\ Fluid Mech. {\bf220}, 1 (1990).}

\bibitem{Cambon97}
{C. Cambon, N.N. Mansour, and F.S. Godeferd, J. Fluid Mech. {\bf 337}, 303 (1997).}

\bibitem{Biferale03}
{L. Biferale {\it et al}, Phys. Fluids {\bf 15}, 2105 (2003).}

\bibitem{Kraichnan73}
{R.H. Kraichnan, J. Fluid Mech. {\bf 59}, 745 (1973).}

\bibitem{Morinishi01}
{Y. Morinishi, K. Nakabayashi, and S. Ren, JSME Int. J. Ser. B {\bf44}, 410 (2001).}

\bibitem{Lesieur}
{J.C. Andr\'e and M. Lesieur, J. Fluid Mech. {\bf 81}, 187 (1977).}

\bibitem{group3}
{P.A. Davidson, P.J. Staplehurst, and S.B. Dalziel, J. Fluid Mech. {\bf 557}, 135 (2006); C. Morize and F. Moisy, Phys. Fluids {\bf 18}, 065107 (2006).}

\bibitem{bellet}
{F. Bellet {\it et al}, J. Fluid Mech. {\bf 562}, 83 (2006).}

%\bibitem{group4}
%{P.D. Mininni and A. Pouquet, Phys. Rev. Lett. {\bf 99}, 254502 (2007);}

\bibitem{Mininni07}
{P.D. Mininni and A. Pouquet, Phys.\ Rev.\ Lett. {\bf 99}, 254502 (2007).}

\bibitem{McWilliams84}
{J.C. Mc{W}illiams, J.\ Fluid Mech. {\bf146}, 21 (1984).}

\bibitem{muller}
{W.C. M\"uller and M. Thiele, Europhy. Lett. {\bf 77}, 34003 (2007);P.D. Mininni, A. Alexakis, and A. Pouquet, Phys. Fluids (in press) arXiv:0802.3714v1 (2008).} 

\bibitem{MininniPress}
{P.D. Mininni and A. Pouquet, (submitted to Phys. Rev. E), arXiv:0707.3620v1 (2008).}

\bibitem{Zhou95}
{Y. Zhou, Phys. Fluids {\bf 7}, 2092 (1995).}

%\bibitem{Mininni08}
%{P.D. Mininni, A. Alexakis, and A. Pouquet, Phys.\ Rev.\ E {\bf77}, 036306 (2008).}

\bibitem{vapor}
{J. Clyne {\it et. al.}, New J. Phys. {\bf 9}, 301 (2007).}

%\bibitem{Chen031}
%{Q. Chen {\bf et. al.}, Phys. Rev. Lett., {\bf 90}, 214503 (2003).}

%\bibitem{Davidson1}
%{P.A. Davidson,P.J. Staplehurst, and S.B. Dalziel, J. Fluid Mech., {\bf 557}, 135 (2006).}

%\bibitem{Davidson2}
%{P.J. Staplehurst,P.A. Davidson, and S.B. Dalziel, J. Fluid Mech., {\bf 598}, 81 (2008).}

%\bibitem{Moisy}
%{F. Moisy,C. Morize and M. Rabaud,Conference on turbulence and interactions TI2006}

%\bibitem{Bellet}
%{F. Bellet, {\bf et al}, J. Fluid Mech., {\bf 562}, 83 (2008).}

%\bibitem{Brachet83}
%{M. Brachet, {\bf et al}, J. Fluid Mech., {\bf 130}, 411 (1983).}

%\bibitem{Mininni07}
%{P.~D. Mininni and A. Pouquet,Phys.\ Rev.\ Lett., {\bf99}, 254502 (2007).}

%\bibitem{Mininni08}
%{P.~D. Mininni, A. Alexakis ,and A. Pouquet,Phys.\ Rev.\ E, {\bf77}, 036306 (2008).}

%\bibitem{Simand00}
%{C.Simand, F.Chill\`a, and J.-F.Pinton, Europhys.\ Lett.,\ {\bf49}, 336 (2000).}

%\bibitem{McWilliams84}
%{J.~C. Mc{W}illiams, J.\ Fluid Mech., {\bf146}, 21 (1984).}

%\bibitem{Zhou95}
%{Y. Zhou, Phys. Fluids {\bf 7}, 2092 (1995).}

%\bibitem{Kraichnan73}
%{R.~H. Kraichnan and J. Fluid Mech., {\bf 59}, 745 (1973).}

%\bibitem{Chen03}
%{Q. Chen, S. Chen, and G.L. Eyink, Physics of fluids, {\bf 15}, 361 (2003).}

%\bibitem{Borue97}
%{V. Borue and S.A. Orszag, J.Phys.Rev.E, {\bf 55}, 7005 (1997).}

%\bibitem{Mininni06}
%{P.~D. Mininni, A. Alexakis, and A. Pouquet,Phys.\ Rev.\ E {\bf74}, 016303 (2006).}

\end{thebibliography}

\end{document}